# IMPROVED MAGNETRON STABILITY AND REDUCED NOISE IN EFFICIENT TRANSMITTERS FOR SUPERCONDUCTING ACCELERATORS*

G. Kazakevich[#], R. Johnson, Muons, Inc, Batavia, IL 60510, USA
V. Lebedev, V. Yakovlev, Fermilab, Batavia, IL 60510, USA

*Abstract*

State of the art high-current superconducting accelerators require efficient RF sources with a fast dynamic phase and power control. This allows for compensation of the phase and amplitude deviations of the accelerating voltage in the Superconducting RF (SRF) cavities caused by microphonics, etc. Efficient magnetron transmitters with fast phase and power control are attractive RF sources for this application. They are more cost effective than traditional RF sources such as klystrons, IOTs and solid-state amplifiers used with large scale accelerator projects. However, unlike traditional RF sources, controlled magnetrons operate as forced oscillators. Study of the impact of the controlling signal on magnetron stability, noise and efficiency is therefore important. This paper discusses experiments with 2.45 GHz, 1 kW tubes and verifies our analytical model which is based on the charge drift approximation.

## INTRODUCTION

Magnetrons are highly-efficient and inexpensive RF generators which allow dynamic fast control of the phase and power [1-3]. The control is necessary to power the SRF cavities, it reduces deviations of the amplitude and phase of the accelerating voltage to levels much less than 1% and 1 degree, respectively, this is required for modern accelerators. Utilization of controlled magnetrons for high-current superconducting accelerators will minimize the capital and operating costs of large-scale projects [4, 5]. Presently known vector methods of fast phase and power management driven by resonant RF signal magnetrons [1, 2], utilize phase-modulated driving signals to control both, phase and power in a wide band. A managed distribution of the total magnetron power between the SRF cavity and a dummy load is used for vector control.

The method described in ref. [3] provides a wideband phase control via the phase-modulated injected resonant signal. The power control is realized by a wide-range magnetron current management. Unlike the vector methods of power control this technique provides the highest average efficiency of the tube at a wide range (up to 10 dB) of power control [3]. The bandwidth of the power control in this case is determined by the bandwidth of the current feedback loop in the magnetron HV power supply. Presently the bandwidth may be ~10 kHz without compromising the efficiency of the HV power supply. This technique is preferable for high-current accelerators.

## BANDWIDTH OF POWER CONTROL REQUIRED FOR SRF ACCELERATORS

The bandwidth of the power control required to suppress "microphonics" can be estimated from the diagram, Fig. 1. Here the dependence of the admissible accelerated (the loading) current (curve C) and the feedback loop bandwidth, δf, required to suppress the "microphonics" to level of 0.3% (curve D) on the loaded Q-factor, $Q_L$, are plotted. Curve C shows the dependence of the loading beam current in a 650 MHz SRF cavity with $R/Q$=610 Ohms and the electric field of 19.9 MV/m on $Q_L$. Graph D shows that for the bandwidth of the feedback loop ≤4.0 kHz the compensation of "microphonics" to level ≈ 0.3% should be at the loading current ≥10 mA.

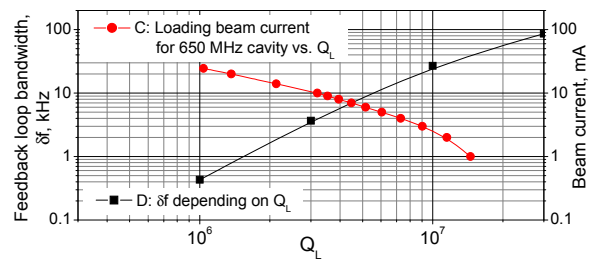

Figure 1: C- the loading beam current for the 650 MHz SRF cavity vs. $Q_L$, D- required bandwidth of the closed feedback loop of power control in SRF cavity vs. its $Q_L$.

The curves were plotted assuming the frequency cutoff of microphonics is ≈ 65 Hz. They demonstrate the advantages of efficient magnetron transmitters that use a wide-range current control for ADS-class projects. The impact of the RF resonant injected signal on stability, noise and efficiency of magnetrons in accordance with the analytical model of magnetron operation is discussed below.

## THE RESONANT INTERACTION OF ELECTRONS WITH A SYNCHRONOUS WAVE IN MAGNETRONS

The resonant interaction of electrons with a synchronous wave in magnetrons was considered by an analytical model [6]. In accordance with the model, the strengths of the electric field of the synchronous wave rotating in the magnetron space of interaction determines the necessary and sufficient conditions for coherent generation of the tube below and above the threshold of self-excitation. The model allows a wide range and fast control of the of the magnetron current (power). The resonant interaction in a conventional magnetron was modelled by a system of equations describing the drift of charges of the Larmor electrons in superposition of the crossed static fields with the RF electric field of the resonant rotating synchronous wave, ref. [3]. The solution of these equations illustrates the phase grouping of the drifting charge into "spokes". The grouped charges provide the energy exchange with the synchronous wave that contributes to coherent generation. This is the basic principle of stable operation of magnetrons. A reduction of the coherent contribution of the drifting charges increases the magnetron's noise and may lead to disruption and damping of the coherent generation. Reference [3] shows that an increase of the synchronous wave amplitude by the resonant injected signal of -10 dB provides the coherent contribution of the drifting charges to the synchronous wave even at low magnetron power. This allows stable generation in the range of the magnetron power over 10 dB at low noise and at the highest efficiency. This is the basis of the technique of power control in magnetrons described in [3].

___________________
* Supported by Fermi Research Alliance, LLC under Contract No. De-AC02- 07CH11359 with the United States DOE in collaboration with Muons, Inc.
[#]e-mail: gkazakevitch@yahoo.com; grigory@muonsinc.com

## EXPERIMENTS VERIFYING THE ANALYTICAL MODEL IN MAGNETRONS OPERATION

The experiments proving the analytical model were performed with 2.45 GHz, 1 kW magnetrons operating in pulsed and CW regimes. The first measurements in pulsed regime demonstrated stable generation of the tubes below and above the threshold of self-excitation. This was studied at the extended range of power control when the magnetrons were driven by a resonant injected signal with power up to 50 W [3]. Detailed measurements were performed in the CW regime with 2.45 GHz, 1.2 kW magnetron type 2M137-IL, Fig. 2, [ibid.].

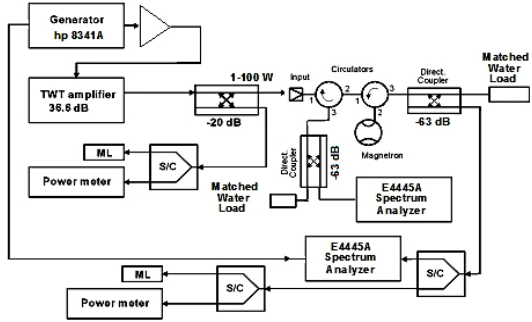

Figure 2: The magnetron setup to test the magnetron frequency stability, noise, and efficiency in CW regime.

The measured offset of the carrier frequency plotted in Fig. 3 shows precise stability at the power range of 10 dB.

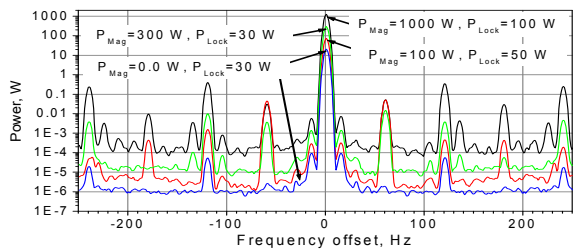

Figure 3: Offset of the carrier frequency at various powers of the magnetron, $P_{Mag}$, and the injected signal, $P_{Lock}$. The trace $P_{Mag}$ =0.0 W, $P_{Lock}$ =30 W shows the frequency offset of the injected signal when the magnetron HV is OFF.

Figure 3 does not demonstrate broadening or shifts of the magnetron spectral line at the range of the magnetron power control of 10 dB. This indicates no notable noise of the magnetron when operating below the threshold of self-excitation and indicates no losses of coherency. This occurs if a sufficient driving resonant signal is injected.

The measurements of noise were performed at the resonant injected signal up to 100 of W. This allows the magnetron output power control in the range over 10 dB.

The power of the coherent synchronous wave induced by the drifting charges is proportional to the square of the charge reaching the magnetron anode. The relative fluctuations of the field in the synchronous wave and in the entire magnetron RF system are caused by fluctuations of the charge reaching the magnetron anode. An increase of the injected resonant signal decreases the charge fluctuations, especially since the amplitude of the synchronous wave becomes more stable at sufficient amplitude of the stable injected signal. Fig. 4 demonstrates the decrease of the measured spectral density of the magnetron noise on power of the injected resonant signal at the nominal power of the tube [6].

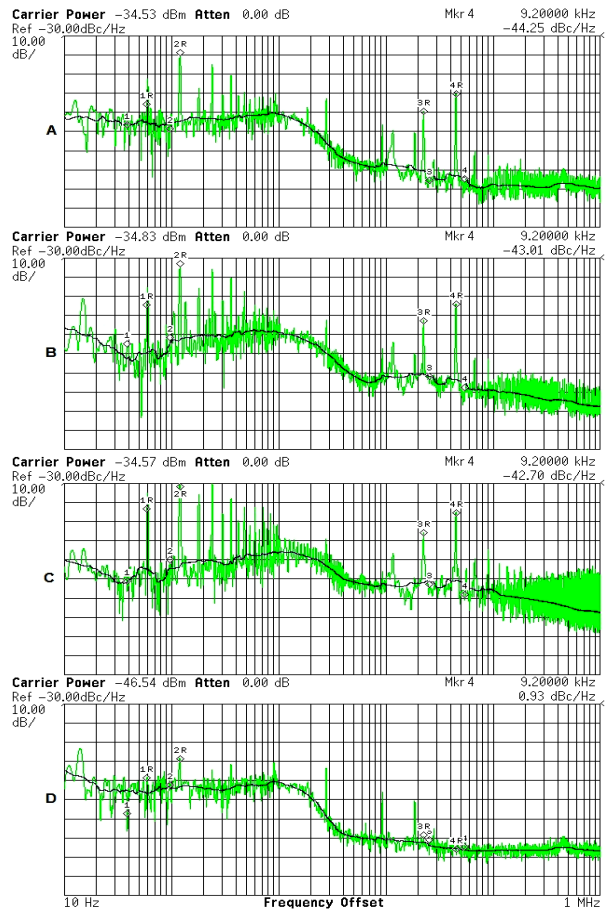

Figure 4: The spectral power density of the magnetron noise at the output power of 1 kW, at the power of the injected signal of 100, 30 and 10 W, traces A, B and C, respectively, [3]. Traces D are the spectral power density of noise of the injected signal with power of 100 W, when the magnetron feeding voltage is OFF. Black traces show the averaged spectral power density of the noise.

If the magnetron operates below the threshold of self-excitation and the injected signal is insufficient for a proper contribution to coherent generation at the resonant energy exchange, this causes damping of the synchronous wave. Damping of the coherent generation results in a dramatic increase of the spectral density of low-frequency noise because of partial (trace B), or total loss of coherency (trace C), respectively, Fig. 5, [6].

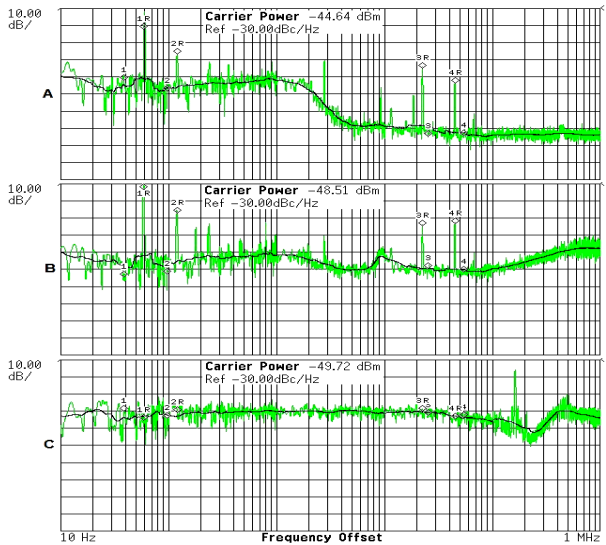

Figure 5: The spectral power density of the noise at various power levels of the locking signal at $P_{Mag}$ = 100 W. A - $P_{Lock}$ =100 W, B - $P_{Lock}$ =30 W, C - $P_{Lock}$ =10 W, Black traces show the averaged spectral density of the noise.

The plots in Fig 4 show that the notable fluctuations of the drifting charges contributing to the coherent generation causes fluctuations of the RF field in the magnetron entire RF system. This results in noise in the MHz range when the magnetron operates at the nominal power. The noise is effectively suppressed by increasing the injection-locking signal to -10 dB.

At a magnetron voltage somewhat less than the threshold of self-excitation, the tube power is rather low, and the induced fields of the synchronous wave are insufficient. This results in loss of coherency of generation at an insufficient injected resonant signal. This causes a dramatic increase of noise in a low-frequency range at a significant decrease of the coherent contribution of the drifting charges into the synchronous wave. Traces B, and C, in Fig. 5 show significant loss of coherency and damping of the coherent oscillation, respectively, [6].

At the injected resonant signal of -10 dB the magnetron demonstrates the noise power density less than -100 dBc/Hz in the frequency range of 1 MHz, avoiding loss of coherency for the output power ranging from 100 W to 1000 W, Figs. 4 and 5, traces A.

Thus optimization of the injected resonant signal allows minimizing the magnetron noise and the tube instability.

The magnetron absolute efficiency, η, was determined as the ratio of the measured magnetron RF power to the output power of the magnetron HV power supply. This was measured at various values of the magnetron RF power [3]. Figure 6 shows the dependence of the magnetron average efficiency <η> as function of the power variation range. Curve D shows results for variations of the magnetron current while curve E relates to the vector methods of power control. The plots demonstrate much higher average efficiency of the magnetron when the RF power is varied in a wide range via current control compared to the vector methods distributing the magnetron power between the SRF cavity and a dummy load.

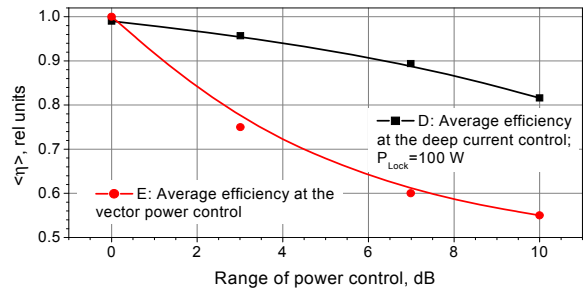

Figure 6: Relative averaged magnetron efficiency vs. range of power control for various methods of control. D- is the average efficiency of the 1.2 kW magnetron driven by the injected resonant signal of -10 dB and measured at deep magnetron current control, [3]. E- is the average efficiency of 1 kW magnetrons with vector power control [1, 2].

## SUMMARY

Basing on the developed analytical model of the resonant interaction of the drifting charge with the synchronous wave in magnetrons was proposed, substantiated and demonstrated method of the magnetron control improving the generation stability, efficiency and reducing the magnetron noise. The method utilizes a relative large injected resonant signal (up to -10 dB of the nominal magnetron power). Experiments with 2.45 GHz, 1 kW magnetrons demonstrated capability of magnetrons for a wide range (over 10 dB) fast power control at highest average efficiency, precise stability and low noise that corresponds to requirements of high-current superconducting accelerators.